\documentclass[aps,prl,showpacs,nobibnotes,twocolumn,
nofootinbib,nobalancelastpage,superscriptaddress,preprintnumbers]{revtex4}
\usepackage{graphicx}
\usepackage{amsmath}
\usepackage{epstopdf}
\usepackage{latexsym}
\usepackage{dcolumn}
\usepackage{bm}
\input{epsf}
\newcommand{\be}{\begin{equation}}
\newcommand{\ee}{\end{equation}}
\newcommand{\ba}{\begin{eqnarray}}
\newcommand{\ea}{\end{eqnarray}}
\newcommand{\bi}{\begin{itemize}}
\newcommand{\ei}{\end{itemize}}
\newcommand{\ga}{\gtrsim}
\newcommand{\bfi}{\begin{figure}
\epsfxsize=9cm
\epsffile}
\newcommand{\efi}{\end{figure}}
\newcommand{\la}{\lesssim}

\newcommand{\gpch}{h^{-1} {\rm Gpc}}
\newcommand{\muk}{\mu {\rm K}}

\newcommand{\mnras}{MNRAS}
\newcommand{\apjl}{ApJ}

\newcommand{\aj}{AJ}
\newcommand{\aap}{AAP}
\preprint{FERMILAB-PUB-10-373-A}
\begin{document}
\title{Confirmation of the Copernican principle at Gpc radial scale and above from
the kinetic Sunyaev Zel'dovich effect power spectrum} 
\author{Pengjie Zhang}
\affiliation{Key Laboratory for Research in Galaxies and Cosmology, Shanghai       
  Astronomical Observatory, Nandan Road 80, Shanghai, 200030,
  China; pjzhang@shao.ac.cn}
\author{Albert Stebbins}
\affiliation{Fermilab Theoretical Astrophysics, Box 500, Batavia, IL
  60510; stebbins@fnal.gov} 
\begin{abstract}
The Copernican principle, a cornerstone of modern cosmology, remains largely
unproven at Gpc radial scale and above.  Here we will show that, violations of
this type will inevitably cause  a first order  anisotropic kinetic Sunyaev
Zel'dovich (kSZ) effect. If  large scale radial inhomogeneities have amplitude
large enough  to explain  the  ``dark energy'' phenomena, the induced kSZ
power spectrum will be much larger than  the ACT/SPT upper limit.  
This single test  confirms  the Copernican principle and rules out the adiabatic
void model as a viable alternative to dark energy. 
\end{abstract}
\maketitle

{\bf Introduction}.---The Copernican principle has been a fundamental tenet of
modern science 
since the 16$^{th}$ century and is also a cornerstone of modern cosmology. It
states that   we should not live in a special region of the universe. Cosmic microwave
background (CMB) observations verify the statistical homogeneity of the last
scattering surface \cite{Hajian06}. Galaxy surveys verify the radial
homogeneity  up to the Gpc scale \cite{Hogg05}.  However, radial homogeneity at
larger scales remains unproven.  

Testing the Copernican principle is of crucial importance for fundamental
cosmology.  If the Copernican principle is violated such that we live in or
near the center of a large ($\sim$ Gpc) void as described by a
Lemaître-Tolman-Bondi (LTB) space-time \cite{Bondi47} in which the matter
distribution is spherically symmetric, the apparent cosmic acceleration
\cite{Riess98,Perlmutter99} can be explained without cosmological constant,
dark energy or modifications of general relativity \cite{LTB}. Throughout the
paper we will restrict to this type of violation of  the Copernican principle.
Various tests of the Copernican principle have been proposed  and  a large
class of void models has been ruled out (e.g. \cite{Goodman95,Caldwell08,GH08}).  Here we propose a powerful single test which confirms the Copernican
principle at Gpc radial scale.

{\bf The kSZ test}.--- A generic
consequence of violating the Copernican principle is that some regions will
expand faster or slower than others and as photons transit between these
regions there will be a relative motion between the average matter frame and
CMB.  When relative motions between free electrons and
photons exist the inverse Compton scattering will induce a shift of the brightness
temperature of CMB 
photons via the kinetic Sunyaev Zel'dovich (kSZ) effect \cite{kSZ}.
This temperature shift will be anisotropic on our sky tracing the anisotropy
of the projected free electron surface density.  This test of the Copernican
principle has been applied to  cluster kSZ  observations
\cite{Goodman95,GH08}, where the electron surface density is high. 
However this effect applies to all free electrons which exist in great
abundance everywhere 
in the universe up to the reionization epoch at redshift $z\ga 6$ (and
comoving distance $\ga 6\gpch$), whereas
clusters are rare  above $z\sim 1$.  So one can expect a more sensitive test
from blank field CMB anisotropy 
power spectrum measurements than from cluster measurements as has been
demonstrated 
for the "dark flow" \cite{Kashlinsky10} induced small scale kSZ effect \cite{Zhang10}.  

Free electrons have local motion $\vec{v}_L$ with respect to the average
matter frame and the subscript ``L''  
refers to ``local''. It vanishes when averaging over sufficiently large scale.
However, when the Copernican principle is violated at large scale, electrons will  have  relative  
motion $\vec{v}_{\rm H}$  between the average matter frame and CMB, which 
does not vanish even when averaging over the Hubble scale. Correspondingly the
induced kSZ temperature 
fluctuation \cite{kSZ,Zhang10}  has two contributions, 
\ba
\label{eqn:ksz}
\Delta T(\hat{n})&=&\Delta T_L(\hat{n})+\Delta T_{\rm H}(\hat{n})\ .
\ea
The first term on the r.h.s is the conventional kSZ effect,
\ba
\Delta T_L(\hat{n})=
T_{\rm CMB}\times \int [1+\delta_{\rm
    e}(\hat{n},z)]\frac{\vec{v}_L(\hat{n},z)\cdot \hat{n}}{c} d\tau_e\ .
\ea
Here, $\hat{n}$ is the radial direction on the sky.  $\tau_{\rm e}$ is the
mean Thomson optical depth to the corresponding redshift and $\delta_{\rm e}$
is the fractional fluctuation in the free electron number density. 
The last term in
Eq. \ref{eqn:ksz} is new and does not vanish in  a non-Copernican universe,
\ba
\label{eqn:CP}
\Delta T_{\rm H}(\hat{n})&=&T_{\rm CMB}\times \int [1+\delta_{\rm
    e}(\hat{n},z)]\frac{\vec{v}_{\rm H}(\hat{n},z)\cdot \hat{n}}{c} d\tau_e
\nonumber \\
&=& 9.1 \mu {\rm K} \left[\int \frac{\vec{v}_{\rm H}\cdot\hat{n}}{10^4 {\rm km/s}}
\frac{\delta_{\rm e}(\hat{n},z)}{0.1}\frac{d\tau_e}{0.001} \right] \ . 
\ea
The last expression neglects the $\int \vec{v}_{\rm H}\cdot \hat{n}d\tau_e$
term,  which has no direction dependence in LTB models in which we live at the 
center,  and is therefore not observable. 
$\vec{v}_{\rm H}$  varies slowly along radial direction and does not suffer
the cancellation of $\vec{v}_L$ in the conventional kSZ effect \cite{OstrikerVishniac86,Vishniac87}. The small scale
anisotropy power spectrum will be quadratic in the amplitude of $\delta_{\rm
  e}$ (which does fluctuate about zero) so we can say that $\Delta T_{\rm H}/T$ is
first order in the density fluctuations. Throughout this paper, unless
otherwise specified, we will focus on this {\it linear} kSZ effect.
We restrict ourselves to adiabatic voids in which the
initial matter, radiation, and baryon densities track each other.  This is
what one would expect if baryogenesis and dark matter decoupling occurs after
the process which generates the void inhomogeneity.    We also restrict ourselves to
voids  outside of which both matter and radiation are homogeneous.
Adding additional inhomogeneities will generically lead to larger values of
$v_H$. 

To explain  the dimming of SNe-Ia and hence the apparent cosmic acceleration
without dark energy and modifications of general relativity,
we shall live in an underdense region (void) of size $\ga 1\gpch$, with a
typical outward velocity $v_{\rm H}\ga 10^4$ km/$s$ (e.g. \cite{GH08}). Given the baryon
density $\Omega_{\rm  b}h^2=0.02\pm 0.002$ from the big bang
nucleosynthesis \cite{Burles01}, $\tau_{\rm e}>10^{-3}$. Scaling the observed weak
lensing rms convergence $\kappa\sim 10^{-2}$ at $\sim 7^{'}$ \cite{lensing},  the rms  fluctuation in $\delta_{\rm e}$ projected over Gpc
length is $\ga 0.1$ at such scale. Hence such a void generates a
kSZ power spectrum $\Delta T_{\rm H}^2\ga 80 \mu \rm  k^2$ at multipole $\ell=3000$.  This is in conflict with recent kSZ observations. The South Pole
telescope (SPT)  collaboration \cite{SPT}  found $\Delta T^2<6.5\muk^2$
($95\%$ upper limit)  and the 
Atacama cosmology telescope (ACT) collaboration \cite{ACT} found
$\Delta T^2<8\muk^2$. This 
simple order of magnitude estimation demonstrates the  potential
discriminating power of 
the kSZ power spectrum measurement. It suggests that a wide range of  void
models capable of replacing 
dark energy are ruled out.  This also demonstrates how purely empirical
measurements of   
CMB anisotropies and the large scale structure (e.g. weak lensing) can in
principle be combined to limit non-Copernican models without any assumptions
of how the inhomogeneities vary with distance. 

We perform quantitative calculation for a popular void model, namely the Hubble 
bubble model (\cite{Caldwell08} and references therein). In this model, we live at the center of a Hubble bubble of constant  matter density $\Omega_0<1$ embedded in a flat Einstein-de Sitter
universe ($\Omega_m=1$).  The void extends to redshift $z_{\rm edge}$,
surrounded by a compensating shell ($z_{\rm edge}<z<z_{\rm out}$) and then the
flat Einstein-de Sitter universe ($z>z_{\rm out}$).  The kSZ effect in
this universe has two components,  
(1) the linear kSZ arising from the large angular scale anisotropies
generated by matter  (a) inside the void, (b) in the compensating  shell,  
(c) outside the void; (2) the conventional  kSZ
effect quadratic in density fluctuation \cite{Vishniac87} and the kSZ effect from
patchy reionization \cite{Santos03}. The contributions of each of these to the
anisotropy power  spectrum are uncorrelated.  Hence the ACT/SPT measurements
put an upper limit on the total.  
The later contributes $\sim 3.5 \muk^2$ \cite{Zhang04},  so what is
left for the first  component is $\la 3\muk^2$. However, we will test
the Copernican principle in a conservative way, by requiring the power
spectrum of the first component generated by matter {\it inside the void} to be
below the SPT upper limit $6.5\muk^2$ at $\ell=3000$. 

For a general Hubble bubble $\vec{v}_{\rm H}$ is determined by both Doppler
and Sachs-Wolfe  anisotropies generated by the void and depends qualitatively
on the size of the void \cite{Sakai93,Caldwell08} .  As we shall see below it is only Hubble
bubbles with $z_{\rm edge}<1$ which are consistent with both the SNe data and
the proposed kSZ test, and for these a simple Doppler formula can be used \cite{Caldwell08,VanAcoleyen08} 
\be
\label{eqn:vH}
v_{\rm H}(z)\approx\left[H_{\rm i}(z)-H_{\rm e}\right]{D_{\rm A,co}(z)\over1+z}
\ee
where, $H_{\rm i}(z)$ is the Hubble expansion rate inside the void as a
function of redshift,  $H_{\rm e}$ gives the Hubble expansion rate exterior to
the void at the same cosmological time, $D_{\rm A,co}(z)$ is the comoving
angular diameter distance to redshift $z$. 

The temperature fluctuation at multipole $\ell$ generated by
the linear kSZ effect inside of the Hubble bubble is, under the Limber
approximation,  
\ba
\label{eqn:cl}
&&\Delta T^2_{\rm H}(\ell)=(9.1\muk)^2
\int_0^{z_{\rm edge}} \left[\frac{v_{\rm H}(z)}{10^4 {\rm km/s}}\right]^2 \\
&&\hskip20pt\times
    \left[\frac{d\tau_e/dz}{0.001}\right]^2
    \left[\frac{\frac{\pi}{\ell} \Delta_{\rm e}^2(\frac{\ell}{D_{\rm A,co}(z)},z)}{0.1^2}\right]
    \frac{D_{\rm A,co}(z)}{c/H_{\rm i}(z)}\,dz \nonumber \ .
\ea
Here $\Delta T_{\rm H}^2\equiv T_{\rm CMB}^2\ell(\ell+1)C_{\ell}/(2\pi)$,
$C_\ell$ is the corresponding angular power spectrum, 
$\Delta_{\rm e}^2(k,z)=\frac{k^3}{2\pi^2} P_{\rm e}(k,z)$ is the dimensionless electron number overdensity at wavenumber $k$ and redshift $z$. 

\bfi{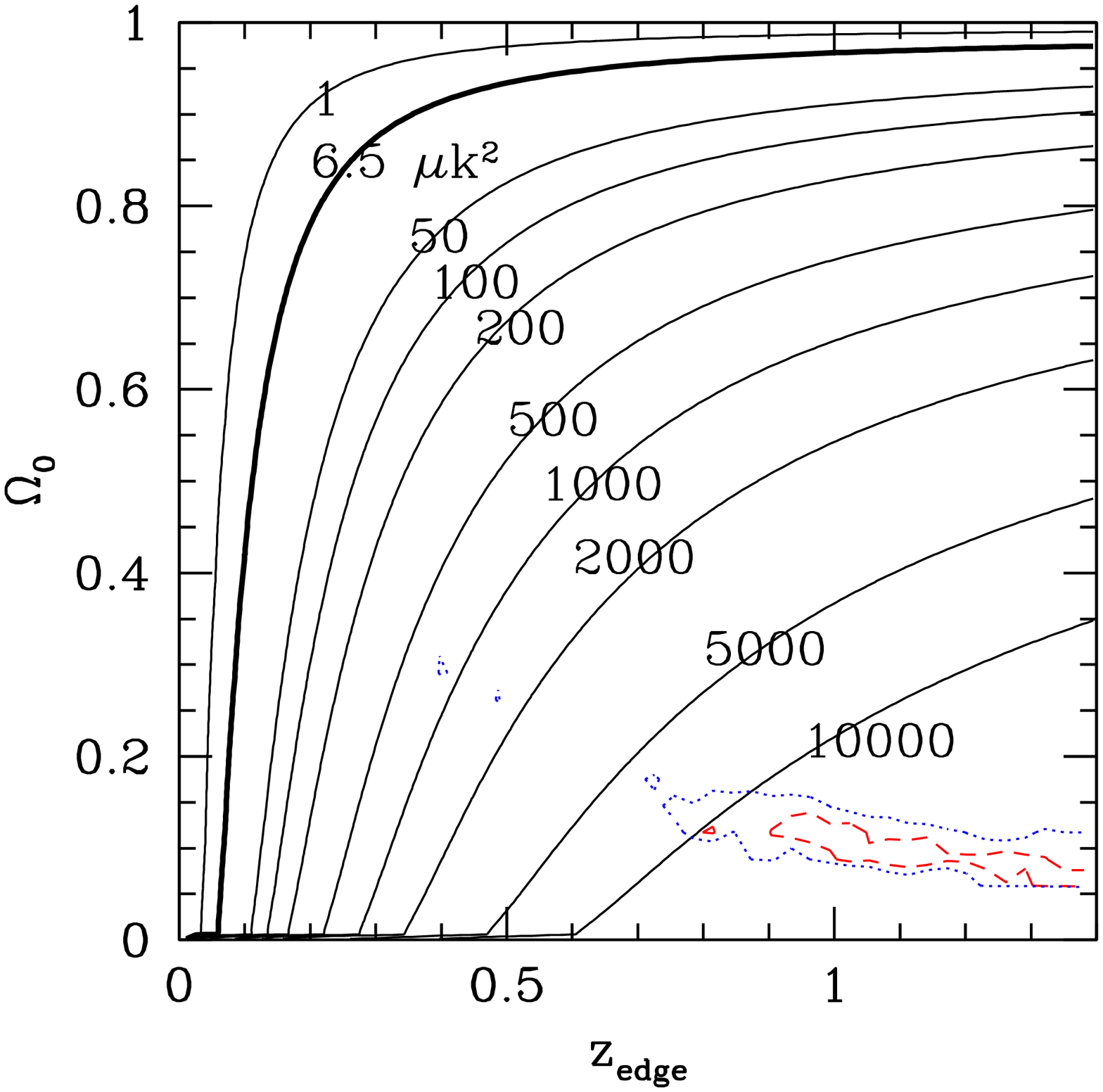}
\caption{The kSZ test.  Black curves have constant $\Delta
  T^2_H(\ell=3000)$. The thick one highlights the SPT $95\%$ upper limit, $\Delta T^2<6.5\muk^2$ \cite{SPT}.  The kSZ test alone  rules out large voids with low density and
strongly supports the Copernican principle. The dashed and dotted contours are the $2$-$\sigma$ and
$3$-$\sigma$ constraints from the UNION2 supernova data \cite{UNION2}. The kSZ
test robustly excludes the Hubble bubble model as a viable alternative to dark
energy.  \label{fig:cl}}  
\efi

\bfi{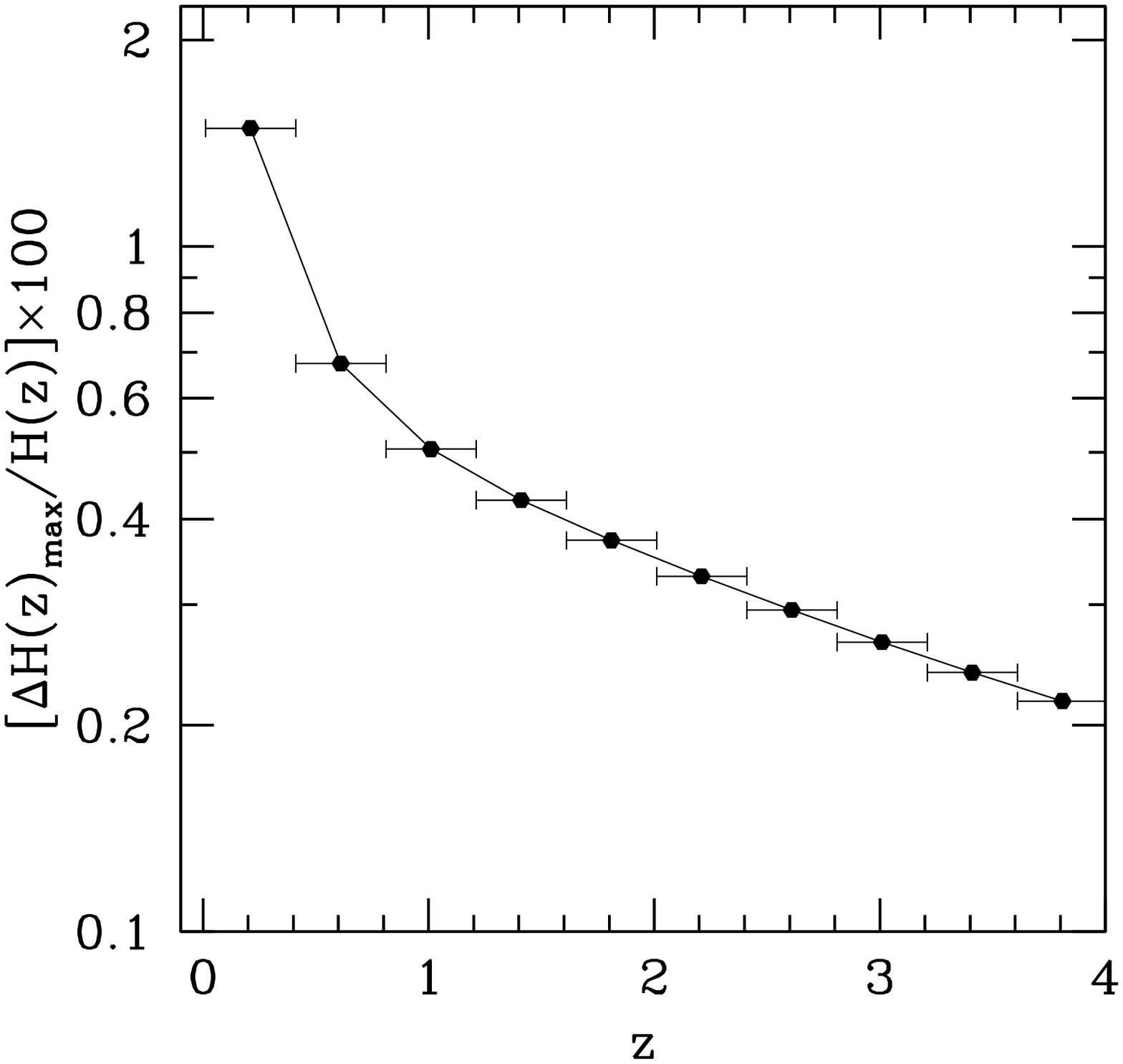}
\caption{The maximal deviation from the overall expansion allowed by the SPT
observation,  for each mass shell of $\Delta z=0.4$, which corresponds to
 $1\gpch$ at  $z\sim 0$,   $0.7 \gpch$ at $z\sim 1$ and $0.5\gpch$ at $\sim
2$. 
 \label{fig:H}}  
\efi  

In our calculations we approximate  $P_{\rm e}$ by the matter power spectrum
$P_{\rm m}$ and approximate $P_{\rm m}$ by it's form in a standard
$\Lambda$CDM cosmology.  It is non-trivial to calculate $P_{\rm m}$ in LTB
models, since even at linear scales the expansion rate is locally anisotropic
so the inhomogeneities will have an anisotropic power spectrum (see
\cite{Clarkson09}), and since we are no longer assuming the cosmological
principle one could also expect large scale variations in the initial
inhomogeneities.  The measured matter clustering and its evolution agree with
the standard $\Lambda$CDM cosmology to a factor of $\sim2$ uncertainty up to
$z\sim 1$  
\cite{lensing,redshiftdistortion} , as do the galaxy clustering and
evolution \cite{galaxy}.   A minimalist approach is to simply use the
$\Lambda$CDM predictions since any viable LTB models must be consistent with
this data.    If this assumption are not satisfied then one should be able to
obtain even tighter constraint by considering these extra tests.  Here we use 
$P_{\rm m}$ calculated by the CMBFAST package \cite{CMBFAST},  nonlinear
clustering from the halofit formula \cite{Smith03} all using  assuming
$\Lambda$CDM with $\Omega_{\rm m}=0.27$,  
$\Omega_{\Lambda}=1-\Omega_{\rm m}$, $\Omega_{\rm b}=0.044$, $\sigma_8=0.84$
and $h=0.71$. All other quantities such as $\tau_{\rm e}$ and $v_H$
  are calculated based on the void model with the same $\Omega_{\rm b}$ and
  $H_{\rm i}(z=0)=100 h$ km/$s/$Mpc.   The kSZ power spectrum is then
computed using  Eq. \ref{eqn:cl}. 

{\bf Constraints on the void model}.---The ACT/SPT upper limit rules out large
voids with low density (Fig. \ref{fig:cl}). Only those voids either with
$\Omega_0\rightarrow 1$  ($\Omega_0\ga 0.8$) or $z_{\rm edge}\rightarrow 0$
($z_{\rm edge}\la 0.2$, corresponding to void radius $\la 0.6 h^{-1}$Gpc)
survive this test (Fig. \ref{fig:cl}).  These results agree fairly well with
those of in a more recent paper  (Fig. 6, \cite{Zibin11}), who have used a
more sophisticated treatment.\footnote{However, since \cite{Zibin11} uses a
  different smooth void model, our results are not directly comparable.}

The kSZ test is highly complementary to other tests such as  the
  supernova  test.
 Our SNe Ia constraint
follows  ref.~\cite{Caldwell08} but uses the improved UNION2 data with 557 SNe
Ia \cite{UNION2}. 
Not allowing for additional intrinsic dispersion of the SNe magnitudes we find a minimum
$\chi^2$ is $605.4$.\footnote{Although this indicates a poor fit including systematic errors
and intrinsic magnitude dispersions would improve the fit.}
Hubble bubble models within $3\sigma$ contour have
typical $\Delta T^2_{\rm H}>10^3\muk^2$ at $\ell=3000$, two orders of
magnitude larger than the SPT upper limit $6.5\muk^2$ \cite{SPT}. On the other hand,
Hubble bubble models consistent with the SPT result have $\Delta \chi^2>209$ ($\chi^2>814$) for the SN Ia test and hence fail too. Thus the combination of SN Ia observations with small scale CMB anisotropy apparently rule out all Hubble bubble models.

Our kSZ calculation is based on these assumptions:
1) $\Omega_{\rm b} h^2$ is the same as in the standard BBN analysis, 2) $P_{\rm m}$ is the same as in a $\Lambda$CDM model,
(based on the argument that any viable void model must reproduce the observed
matter clustering),
3) $P_{\rm e}=P_{\rm m}$ (good to $\sim10\%$ accuracy \cite{Jing05}),   
4) eq.~\ref{eqn:vH} for velocities (roughly accurate for sub-horizon
voids\cite{VanAcoleyen08} which is required by CMB data \cite{Caldwell08,GH08}),    
5) neglect of kSZ contributions from the compensating shell (which would only
increase kSZ anisotropy),  
6) a simple adiabatic Hubble bubble void, 
and
7) no CMB flow (intrinsic dipole) from non-adiabatic initial conditions
outside the void.
 We expect that relaxing 1)-6) in reasonable ways could not
  significantly reduce the tension imposed by the kSZ test, since for void models to explain
  the observed SN dimming, they must have  
large scale  gravitational potential of large amplitude and hence must have
large $v_H$ and large kSZ effect. For example, \cite{Zibin11} adopted a void
model of different profile and  found much
weaker SN constraint, but the generated kSZ power is nevertheless
much larger than the ACT/SPT upper limit. This demonstrates the great
discriminating power of the kSZ test.   Completely relaxing 7) could 
change our conclusion for rather contrived initial conditions
\cite{Clarkson10}, but would 
generically lead to even larger and more unacceptable kSZ effect. Thus
comparing kSZ  
with SNe is by far the most stringent  test of the void models and
the Copernican principle at Gpc scale and above.   
We conclude that any adiabatic void models capable of explaining the
supernova Hubble diagram would likely generate too much kSZ power on
the sky to be consistent with the ACT/SPT upper limit. This
strengthens the evidences for cosmic acceleration and dark energy.

{\bf Constraints on the Hubble flow}.---
Still,  violation of the Copernican principle less dramatic than the above
void models may exist \cite{Thomas11}. For example, there could be  large scale
density modulation on the $\Lambda$CDM  background. As long as the amplitude of
the modulation is   
sufficiently small, it can pass the supernova test and the structure growth
rate test. However, if unaccounted, it could bias the dark energy constraint. The kSZ test is able to put interesting constraint on this
type of violation.  We take a model independent approach and parameterize  the
violation of the Copernican  principle by $\Delta H(z)$, the deviation of the Hubble
expansion of a mass shell of size  $\Delta z$ centered at redshift $z$ from
the overall expansion of the background universe.  The
ACT result  constrains $|\Delta H(z)/H(z)|\la 1\%$ for each  mass shell of radial
width $\sim 1\gpch$  (Fig. \ref{fig:H}). This estimation neglect contributions
from other mass shells so the actual constraint is tighter.  This test can be
carried out on each 
patch of the sky  to test the isotropy of the Hubble flow.

The above test is not able to determine at which redshift a violation of the
Copernican principle occurs, since the kSZ power spectrum is the sum over all 
contributions along the line-of-sight and hence has no redshift
information. This problem can be solved with the aid of  a survey of the large
scale structure (LSS) with redshift information. 

The basic idea is the same as the one
proposed by \cite{Zhang10} to probe the dark flow through the kSZ-LSS density
distribution two point cross correlation.  This cross correlation is non-zero only in non-Copernican
Universes, since the velocity $\vec{v}_{\rm H}$ varies slowly over the
clustering length of the LSS and since the linear kSZ effect is linear in
density.  Since the cross correlation vanishes for the conventional kSZ
effect, a non-vanishing cross correlation signal can serve as a smoking gun of
violation of the Copernican principle. The thermal SZ contaminates the measurement. However, it can be largely removed by spectral fitting or
observing at its null: $217$ GHz.  Since the redshift surveys
can map the LSS  with much higher S/N than kSZ measurements,
this cross correlation can achieve much higher S/N than the kSZ auto-correlations.
We thus expect that small scale CMB anisotropy surveys, such as ACT and SPT, in
combination with deep LSS surveys  will be able to put more stringent constraints on
violations of the Copernican principle at each redshift and each direction of
the sky. 

{\bf Acknowledgment}.---
We thank U. Pen, J. Fry, C. Clarkson, S. Das, P. Ferreira and J. Zibin for useful
discussions. PJZ was  
supported by the NSFC grants and the Beyond the Horizons program. A.S was supported by the DOE at
Fermilab under contract No. DE-AC02-07CH11359.


\begin{thebibliography}{1}

\bibitem{Hajian06} A. Hajin \& T. Souradeep, T., \prd, 74, 123521 (2006)

\bibitem{Hogg05}  D.Hogg et al., ApJ, 624, 54 (2005)

\bibitem{Bondi47} H. Bondi, \mnras, 107, 410 (1947)

\bibitem{Riess98}  A. Riess et al., \aj, 116, 1009 (1998)

 \bibitem{Perlmutter99}  S. Perlmutter et al., \apj, 517, 565 (1999)

\bibitem{LTB}   e.g. J. Moffat and D. Tatarski, ApJ, 453, 17(1995); M. Celerier,  Astron.\ Astrophys., 353, 63 (2000);
 R. Barrett, and C. Clarkson, C.A.Class.\ Quant.\ Grav.,17,
  5047 (2000); K. Tomita, K., \mnras, 326, 287 (2001)

\bibitem{Goodman95} J. Goodman, \prd, 52, 1821 (1995); C. Clarkson,
  B. Bassett, and T.~H.-C. Lu, \prl,
  101, 011301 (2008); T. Clifton, P.~G. Ferreira, and K. Land,   \prl, 101,
  131302  (2008); J. Uzan, C. Clarkson and G. Ellis. Phys.Rev.Lett., 100,
  191303 (2008) ;T. Biswas, A. Notari \& W. Valkenburg. JCAP,11, 030 (2010);
  A. Moss, J.P. Zibin \& D.Scott., arXiv:1007.3725 (2010);
  V. Marra \& M. Paakkonen. JCAP,12, 021 (2010)

\bibitem{Caldwell08} R.~R. Caldwell, A. Stebbins,  \prl, 100, 191302 (2008)
 


\bibitem{GH08} J. Garc{\'{\i}}a-Bellido and T. Haugb{\o}lle, JCAP, 9, 16
  (2008); C.Yoo, K. Nakao and  M.Sasak  arXiv:1008.0469 
  (2010) 



\bibitem{kSZ}  R.~A. Sunyaev, and Y.~B. Zeldovich, Comments on
  Astrophysics and Space  Physics, 4, 173 (1972); R.~A. Sunyaev, and Y.~B. Zeldovich,  \mnras, 190, 413 (1980)


\bibitem{Kashlinsky10} A. Kashlinsky, et al. ApJ, 712, L81 (2010)

\bibitem{Zhang10} 
P.J. Zhang, \mnras, 407, L36 (2010). [arXiv:1004.0990]

\bibitem{OstrikerVishniac86} J.P. Ostriker, and E.~T. Vishniac, \apjl, 306,
  L51 (1986)

\bibitem{Vishniac87} E.~T. Vishniac, \apj, 322, 597 (1987)

\bibitem{Burles01} S. Burles, K. Nollett, and M. Turner, \apjl, 552, L1
  (2001)

\bibitem{Clarkson09} C. Clarkson, T. Clifton, S. February. JCAP, 06, 025
  (2009). [arXiv:0903.5040]
  
\bibitem{lensing} L. Fu et al. \aap, 479, 9 (2008); T. Schrabback et
  al. arXiv:0911.0053 (2009) 

\bibitem{SPT} N.~R. Hall, et al., arXiv:0912.4315 (2009); E. Shirokoff, et
  al. arXiv:1012.4788
  
\bibitem{ACT} S.Das, et al. arXiv:1009.0847 (2010); J. Dunkley, et
  al. arXiv:1009.0866 (2010) 

\bibitem{Jing05} Y. Jing, et al. \apj, 640, L119 (2006)

\bibitem{Santos03} M.~G. Santos et al., \apj, 598, 756  (2003)

\bibitem{Zhang04} P.J. Zhang, U.-L. Pen, and H. Trac, \mnras, 347, 1224 (2004) 
  
\bibitem{Sakai93} N. Sakai, K. Maeda, and M. Sasaki, Prog. Theor. Phys., 89, 1193 (1993) 

\bibitem{VanAcoleyen08} K. Van Acoleyen, JCAP, 10, 28 (2008)


\bibitem{redshiftdistortion} L. Guzzo et al., \nat, 451,
  541 (2008); C. Blake et al., arXiv:1003.5721 (2010)  
\bibitem{galaxy} M. Tegmark, et al., \apj, 606, 702 (2004); A.~L. Coil et
  al., \apj, 644, 671 (2006); Z. Zheng, A.~L. Coil, and I. Zehavi, \apj, 667, 760 (2007)


\bibitem{CMBFAST} U. Seljak, M.  Zaldarriaga, \apj, 469, 437 (1996)

\bibitem{Smith03} R.~E. Smith et al., \mnras, 341, 1311 (2003)

\bibitem{Zibin11} J.P. Zibin, A. Moss. 2011, arXiv:1105.090

\bibitem{UNION2} R. Amanullah et al. ApJ, 716, 712 (2010)
\bibitem{Clarkson10} C. Clarkson, M. Regis. arXiv:1007.3443 (2010)
\bibitem{Thomas11} S. Thomas,  F. Abdalla, O. Lahav. \prl, 106, 241301 (2011)



\end{thebibliography}
\end{document}